\documentclass[12pt]{iopart}
\usepackage{iopams}
\usepackage{graphicx}
\usepackage{subeqn}
\usepackage{subfigure}
\newcommand \etc {{\sl etc.\ }}
\newcommand \eg {{\sl e.g.\ }}
\newcommand \ie {{\sl i.e.\ }}
\newcommand \viz {{\sl viz.\ }}
\newcommand \fig[1] {Fig.\ \ref{#1}}

\newcommand \beq {\begin{equation}}
\newcommand \eeq {\end{equation}}
\newcommand \beqa {\begin{eqnarray}}
\newcommand \eeqa {\end{eqnarray}}
\newcommand \bsubeq {\begin{subequations}}
\newcommand \esubeq {\end{subequations}}
\newcommand \benum {\begin{enumerate}}
\newcommand \eenum {\end{enumerate}}
\newcommand \bfig {\begin{figure}[!t]\begin{center}}
\newcommand \efig {\end{center}\end{figure}}
\newcommand \btab {\begin{table}[!ht]\begin{center}}
\newcommand \etab {\end{center}\end{table}}
\newcommand \lrnd {\left(}
\newcommand \rrnd {\right)}
\newcommand \lsq {\left[}
\newcommand \rsq {\right]}

\newcommand \lang {\left\langle}
\newcommand \rang {\right\rangle}
\newcommand \ltvert {\left\vert}
\newcommand \rtvert {\right\vert}

\newcommand \prl[3] {Phys.\ Rev.\ Lett.\ #1,\ #2\ (#3)}
\newcommand \prd[3] {Phys.\ Rev.\ D#1,\ #2\ (#3)}
\newcommand \prc[3] {Phys.\ Rev.\ C#1,\ #2\ (#3)}

\newcommand \plb[3] {Phys.\ Lett.\ B#1,\ #2\ (#3)}

\newcommand \jhep[3] {JHEP\ #1,\ #2\ (#3)}

\begin{document}

\title[]{Fluctuations, correlations and some other recent results from lattice QCD}

\author{Swagato Mukherjee}

\address{Physics Department, Brookhaven National Laboratory, Upton, NY 11973, USA}

\ead{swagato@bnl.gov}

\begin{abstract}
I summarize recent results from lattice QCD on a few selected topics which are of
interest to the heavy-ion physics community. Special emphasis is placed upon
observables related to fluctuations of conserved charges and their connection to
event-by-event fluctuations in heavy-ion collision experiments.
\end{abstract}



\section{Transition temperature of QCD}
\label{sec:Tc}

At high temperatures and/or densities ordinary hadronic matter is expected to undergo
a transition to a new form of matter, \viz the Quark Gluon Plasma (QGP). The
spontaneously broken chiral symmetry of Quantum Chromodynamics (QCD) is also expected
to become restored at this point. The temperature at which such a transition takes
place is generally referred to as the QCD transition temperature ($T_c$). The value
of $T_c$ is one of the most basic input from lattice QCD to the heavy-ion
phenomenology. At present almost all the precision state-of-the-art finite
temperature/density lattice QCD computations are being performed using
computationally cheaper staggered fermion discretization scheme. At any non-zero
lattice spacing ($a$) staggered fermions give rise to a distorted hadron spectrum,
which apart from containing a single Goldstone pion also contains 15 unphysical
pions. Masses of these unphysical pions are larger than the Goldstone pion and the
physical hadron spectrum is obtained only in the continuum limit $a\to0$. Since the
chiral property of the staggered fermions are sensitive to the masses of these
unphysical pions it is essential to reduce their masses (relative to the mass of the
Goldstone pion) while studying the QCD transition. 

One of the major progress made during the last couple of years in reducing this
lattice artifact is the use of Highly Improved Staggered Quark (HISQ) action. The HISQ
action largely reduces the masses of these unwanted pions even for relatively coarser
lattices and consequently facilitates the continuum $a\to0$ extrapolation using
moderate values lattice spacings. In this conference the HotQCD collaboration
presented \cite{bazavov-1} their latest results on $T_c$ obtained from observables
related to the chiral symmetry restoration using two different types of fermion
discretization (\viz HISQ and Asqtad actions) giving consistent results in the
continuum $a\to0$ limit.  HotQCD collaboration has also performed
interpolations/extrapolations of their results to the physical values of quark masses
using $O(N)$ scaling analysis. At the physical quark masses the continuum
extrapolated value for the transition temperature is \cite{bazavov-1,hotqcd}--- $T_c
= 157\pm4\pm3\pm1$ (HotQCD\ preliminary). These latest results from the HotQCD
collaboration is in excellent agreement with the previously published results
($T_c=147-157$ MeV, obtained using observables which are sensitive to the light
quarks) of the Wuppertal-Budapest collaboration \cite{fodor-1,borsanyi-1}.


\section{Universal properties of the chiral transition}
\label{sec:scaling}

\bfig
\subfigure[]{ \label{fig:phase} \includegraphics[width=4.5cm,height=3.5cm]{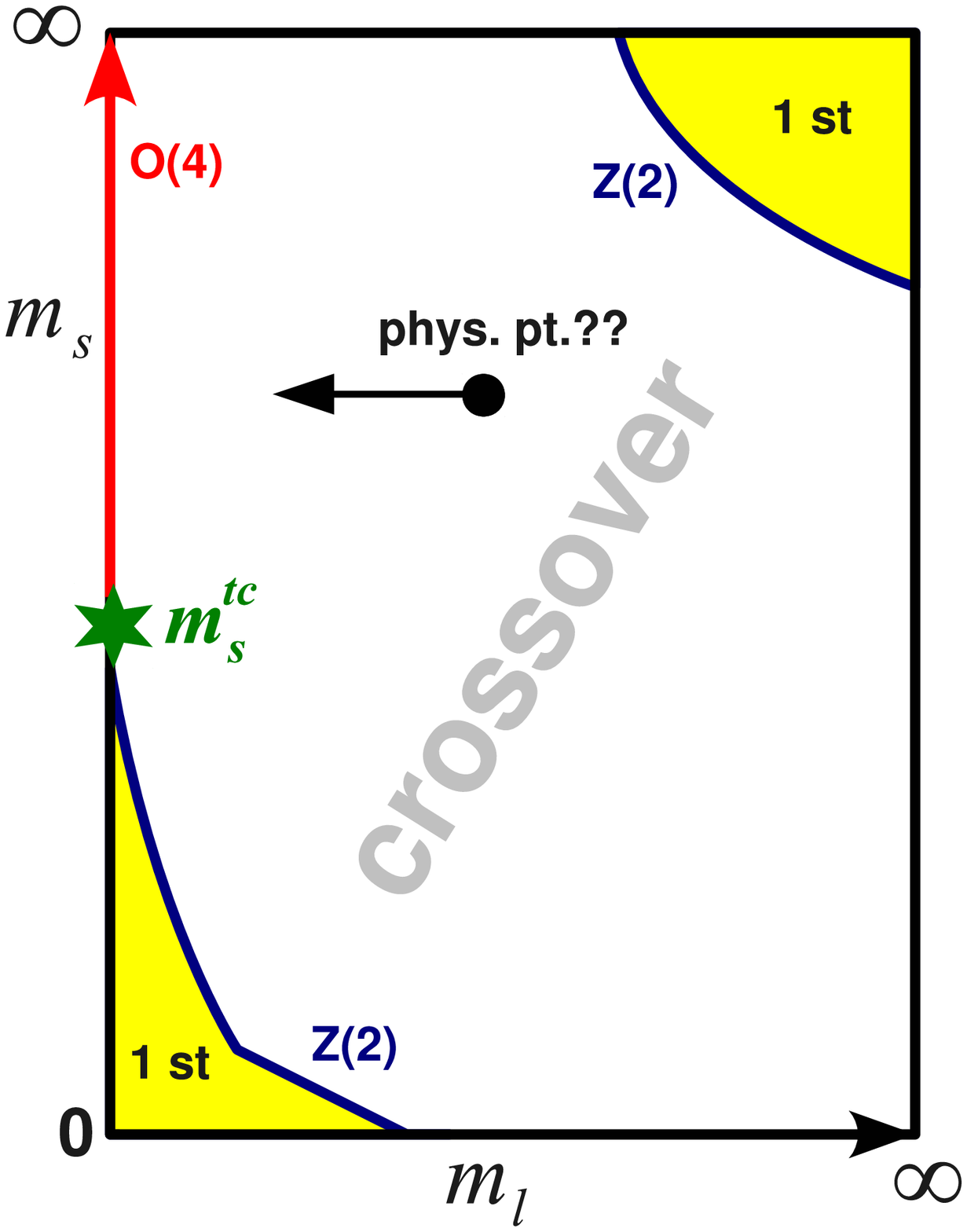} }
\hspace{2cm}
\subfigure[]{ \label{fig:scaling} \includegraphics[scale=0.4]{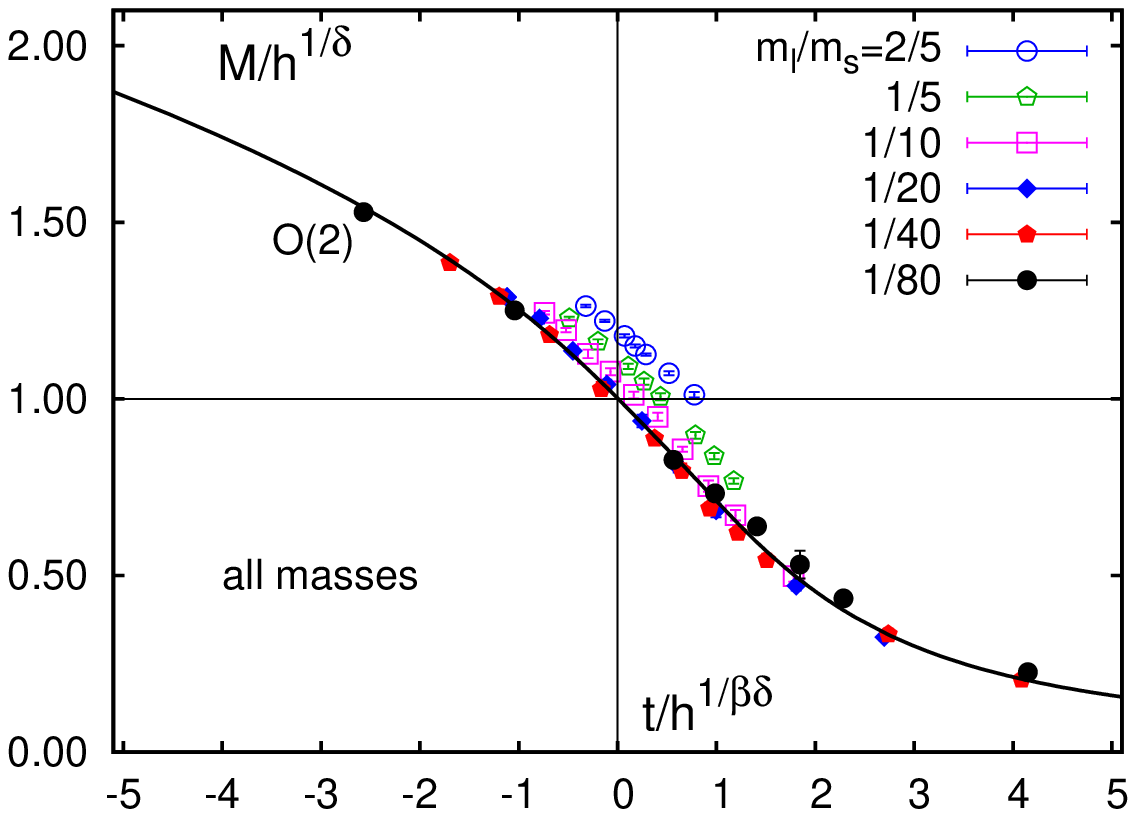} }
\vspace{-0.25cm}
\caption{(a) A sketch depicting the nature of QCD transition as functions of the
light ($m_u=m_d=m_l$) and strange ($m_s$) quark masses at zero baryon chemical
potential. (b) Evidence of $O(N)$ scaling of the order parameter as observed in Ref.\
\cite{ejiri-1}. See text for details.}
\efig

The nature of the QCD transition depends crucially on the values of the quark masses.
As for example, \fig{fig:phase} shows a sketch of the nature of the transition as
functions of the quark masses for a theory with two degenerate light (up and down)
quarks with masses $m_l=m_u=m_d$ and a heavier strange quark with mass $m_s$ at zero
baryon chemical potential. In the limit $m_l\to0$ and $m_s\to\infty$, the relevant
symmetry is isomorphic to the $3$-d $O(4)$ spin model and the transition is second
order belonging to the $3$-d $O(4)$ universality class. For $m_l=m_s\to0$ the
transition is first order. In the intermediate quark mass region a (rapid) crossover
takes place from the hadronic to the QGP phase. The first order region is separated
from the crossover region by a line of second order phase transitions belonging to
the $3$-d Ising $Z(2)$ universality class. The first order region, the second order
$Z(2)$ and the second order $O(4)$ lines meet at a tri-critical point characterized
by a certain value $m_s^{tc}$ of the strange quark mass. While it is well established
that for the physical values of the quark masses $m_l=m_l^{phys}$, $m_s=m_s^{phys}$
the transition is a crossover \cite{aoki-1} the location of the physical point in the
context of \fig{fig:phase} remains an open issue.  More specifically, if
$m_s^{phys}>m_s^{tc}$ then in the limit $m_l\to0$ one should see a second order
transition belonging to the $3$-d $O(4)$ universality class. On the other hand, if
$m_s^{phys}<m_s^{tc}$ then in $m_l\to0$ limit the transition should be first order. 

This question can be addressed by studying the the universal scaling behavior of the
chiral transition as a function of decreasing light quark mass while keeping the
strange quark mass fixed at its physical value. In the vicinity of a second order
phase transition the behavior of the free energy of a system is largely governed by
the universal scaling properties--- $f \lrnd m_l,m_s,T,\mu_B \rrnd = h^{1+1/\delta}
f_s(z) + f_{reg} \lrnd m_l,m_s,T,\mu_B \rrnd$ , where $h = m_l/(h_0m_s)$ and $t=
(T-T_0)/(t_0T_0)$ are, respectively, the explicit (chiral) symmetry breaking field
and the reduced temperature variable. $T_0$ denotes the critical temperature in the
chiral limit ($h\to0$) and $z=th^{-1/\beta\delta}$ is the so-called scaling variable.
The scaling properties of the order parameter ($\sim\partial f/\partial h$) is given
by $M = m_s \lang\bar\Psi\Psi\rang / T^4 = h^{1/\delta} f_G(z)$, where
$\lang\bar\Psi\Psi\rang$ is the chiral condensate.  The critical exponents ($\beta$,
$\delta$) and the scaling functions ($f_s(z)$, $f_G(z)$) uniquely characterize the
universality class of the transition. The parameters $t_0$, $h_0$ and $T_0$ are
non-universal, \ie they depend on the action, lattice spacing, value of the bare
strange quark mass \etc. The function $f_{reg}(m_l,m_s,T,\mu_B)$ is not related to
universal scaling properties and depicts the usual non-critical regular part of the
free energy. 

Recently such scaling studies were performed in Ref.\ \cite{ejiri-1} by the
BNL-Bielefeld collaboration using p4fat3 staggered quark action. These studies show
that in the limit $m_l\to0$, keeping $m_s=m_s^{phys}$, the chiral transition belongs
to $3$-d $O(N)$ universality class  \footnote{As mentioned before in Section
\ref{sec:Tc}, for staggered fermions at non-zero lattice spacings there is only one
Goldstone pion in the chiral limit as opposed to three in the continuum.  Hence, for
staggered fermions the relevant symmetry group is $3$-d $O(2)$ instead of the $3$-d
$O(4)$ of the continuum.}. As for example, \fig{fig:scaling} shows that the properly
scaled order parameter, $h^{-1/\delta}M$, agrees with the relevant $O(N)$ scaling
function $f_G(z)$ for $m_l\lesssim m_s/20$ which corresponds to a Goldstone pion mass
of $m_\pi\simeq150$ MeV. These studies indicate that the $m_s^{phys}$ is probably
larger than $m_s^{tc}$. Furthermore, they also indicate that the physical QCD
($m_l\simeq m_s/27$) may also lie within the scaling regime of the actual chiral
transition.  However, these studies where performed using p4fat3 staggered action
which is known to have large discretization errors. Hence, in order to confirm these
results similar studies have to be performed using the HISQ action. Such studies are
currently underway.


\section{Curvature of the phase transition line for small $\mu_B$}
\label{sec:curvature}

\bfig
\subfigure[]{ \label{fig:phase1} \includegraphics[scale=0.3]{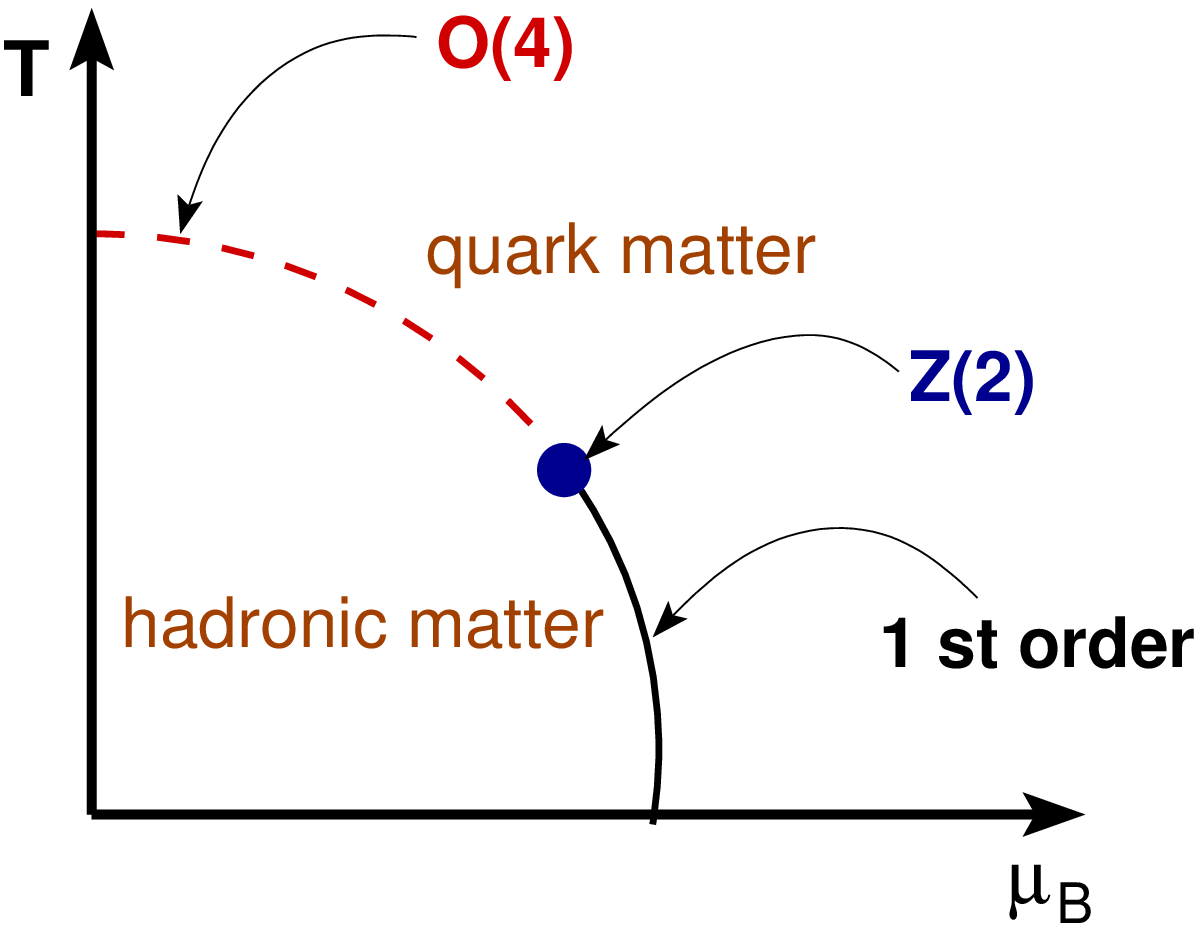} }
\subfigure[]{ \label{fig:kappa} \includegraphics[scale=0.4]{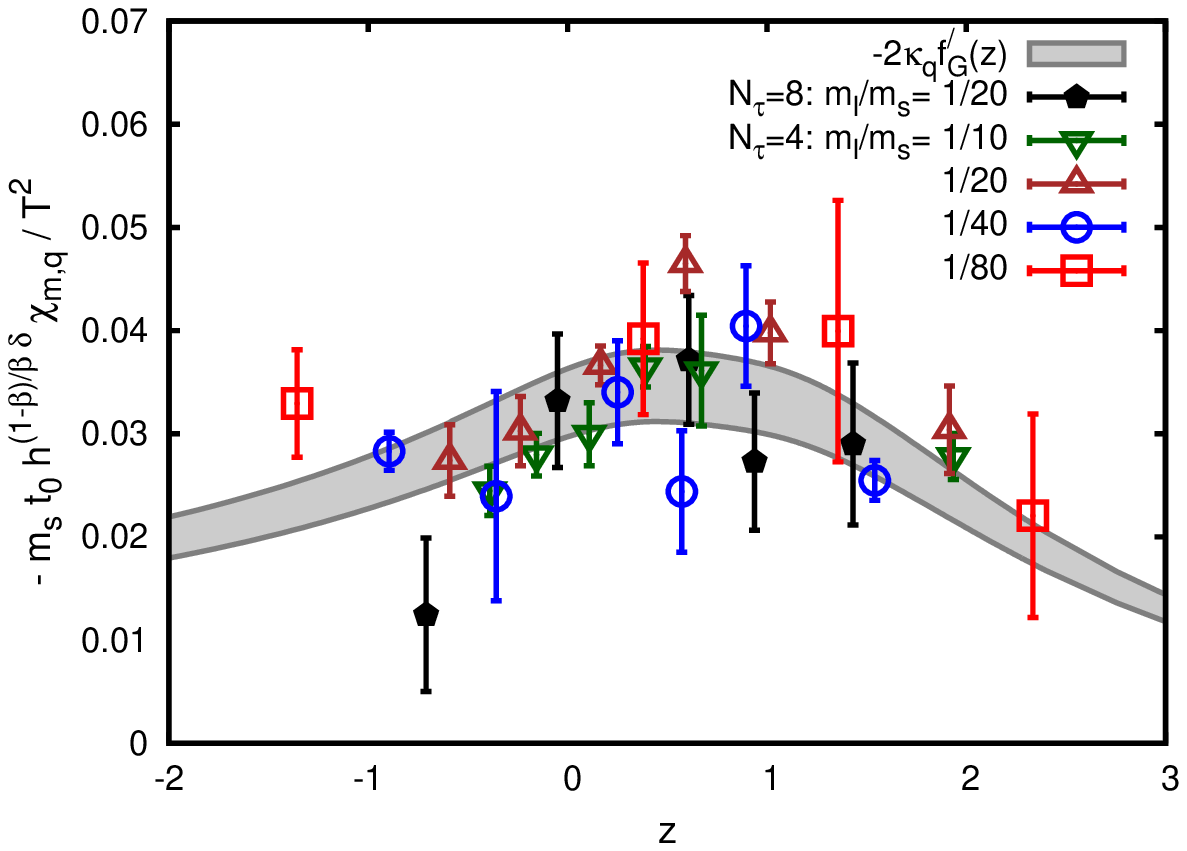} }
\subfigure[]{ \label{fig:curv-fo} \includegraphics[scale=0.4]{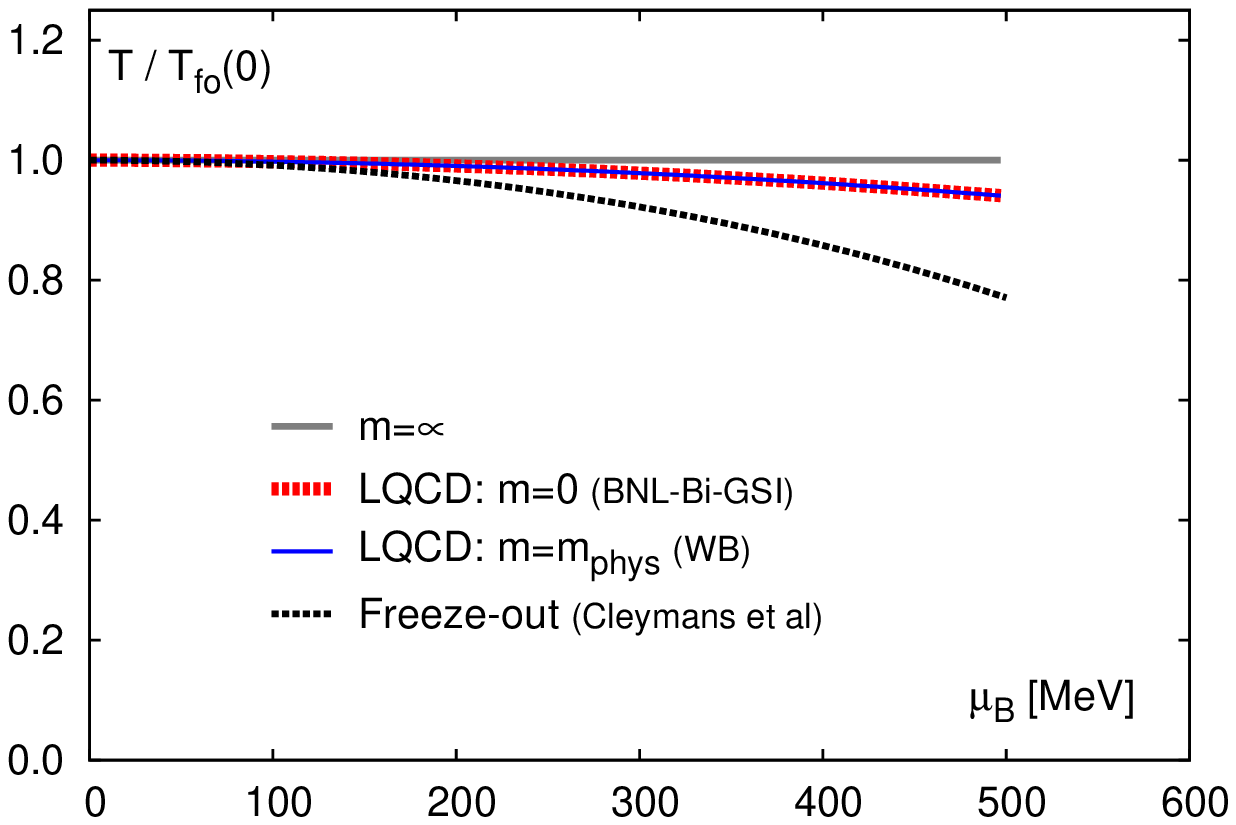} }
\vspace{-0.25cm}
\caption{(a) A sketch of the conjectured phase diagram of QCD in the
temperature--chemical potential plane in the limit of massless light quarks. (b)
Scaling analysis of the ``thermal'' susceptibility of the order parameter and
determination of the curvature of the chiral critical line \cite{kaczmarek-1}. (c)
Comparison of the phase boundary in the $T-\mu_B$ plane and the chemical freeze-out
curve. See text for details.} 
\efig

Based on the indications from the scaling studies mentioned in the previous section
the conjectured phase diagram of QCD in the limit of massless light quarks is
depicted in \fig{fig:phase1}. At zero chemical potential ($\mu_B$) the chiral
transition is second order, belonging to the $3$-d $O(4)$ universality class. This
$O(4)$ chiral phase transition line extends in the $T-\mu_B$ plane and meets the 1st
order phase transition line emanating from the $T=0$ axis at the so-called QCD
(tri-)critical point which belongs to $3$-d $Z(2)$ universality class. The universal
scaling properties of the chiral transition has been extended \cite{kaczmarek-1} to
compute the curvature of the $O(4)$ chiral critical line for small values of baryon
chemical potential $\mu_B/T<<1$. Since the baryon chemical potential does not
explicitly break the chiral symmetry, from the perspective of the universal scaling
properties of the chiral transition, $\mu_B$ is equivalent to the thermal variable.
Further, due to $CP$-symmetry of the free energy, the reduced temperature variable
must be even in $\mu_B$--- $t = \lsq (T-T_0)/T_0 + \kappa_B \lrnd\mu_B/T\rrnd^2 \rsq
/t_0$. Here we assume $\mu_B/T<<1$ and consider only the leading order term in
$\mu_B/T$.  The $O(4)$ chiral transition line in $T-\mu_B$ plane is defined by the
condition $t=0$ and hence at the leading order in $\mu_B$ the chiral transition
temperature is given by--- $T_0(\mu_B)/T_0(0)=1-\kappa_B[\mu_B/T_0(0)]^2$.  Thus
$\kappa_B$ is the curvature of the chiral transition line around $\mu_B=0$.  The
value of the this curvature parameter $\kappa_B$ can be determined by studying the
universal scaling properties of the so-called ``thermal'' susceptibility of the order
parameter--- $\chi_{m,B} = \partial^2 M / \partial(\mu_B/T)^2 = 2\kappa_B
h^{(\beta-1)/\beta\delta} f^\prime_G(z) /t_0$, where $f^\prime_G(z)$ is a scaling
function uniquely associated with the specific universality class. Once the
non-universal scaling parameters $t_0$, $h_0$ and $T_0(0)$ is determined from the
scaling analysis at $\mu_B=0$, as have been done in Ref.\ \cite{ejiri-1}, scaling
analysis of $\chi_{m,B}$ can be preformed in order to determine the only unknown
parameter $\kappa_B$.  

Such a scaling analysis of the ``thermal'' susceptibility of the order parameter was
performed by the BNL-Bielefeld-GSI collaboration in Ref.\ \cite{kaczmarek-1}, using
p4fat3 staggered fermions with two different lattice spacings, and the curvature of
the chiral transition line was determined to be $\kappa_B=0.0066(9)$. In
\fig{fig:kappa} we show the scaling analysis of the ``thermal'' susceptibility of the
order parameter from Ref.\ \cite{kaczmarek-1} \footnote{In Ref.\ \cite{kaczmarek-1}
the ``thermal'' susceptibility of the order parameter $\chi_{m,q}$ was defined with
respect to the average light quark chemical potential $\mu_q=(\mu_u+\mu_d)/2$. Since
for a two flavor theory $\mu_q=\mu_B/3$, the curvature parameter $\kappa_q$ of Ref.\
\cite{kaczmarek-1} is trivially related to the curvature parameter discussed here
through the relation $\kappa_B=\kappa_q/9$.}.  Once the the curvature of the chiral
transition line is known it is possible to compare the phase boundary in $T-\mu_B$
plane, given by $T_0(\mu_B)/T_0(0)=1-\kappa_B[\mu_B/T_0(0)]^2$, with the (chemical)
freeze-out curve \cite{cleymans-1}. This gives an idea about how far the chiral
critical line is from the freeze-out of the heavy-ion collision experiments. Such
comparison is shown in \fig{fig:curv-fo}. In the limit $m_l\to\infty$, \ie for a pure
gluonic theory, the transition temperature is independent of $\mu_B$ and in the limit
$m_l\to0$ the phase boundary is the one obtained by the BNL-Bielefeld-GSI
collaboration in Ref.  \cite{kaczmarek-1}. The transition line for the physical quark
mass should lie in between these two limits. In fact, the phase boundary for the
physical quark masses has also been computed recently from lattice QCD by the
Wuppertal-Budapest collaboration \cite{fodor-1,endrodi-1}. As can be seen from
\fig{fig:curv-fo}, the phase boundary for the physical quark mass almost coincides
with that in the chiral limit and they agrees with the freeze-out curve
\cite{cleymans-1} till $\mu_B\sim150$ MeV and then slowly starts to deviate. However,
one has to remember that the lattice computations for the phase boundary are at
present performed only upto the leading order in $\mu_B$. 


\section{Fluctuation, correlations: lattice QCD meets experiments}
\label{sec:fluctuation}

Although lattice QCD has proven to be a successful technique for performing
ab-initio, quantitative studies of QCD a direct computation at non-zero baryon
chemical potential is, unfortunately, plagued by the infamous sign problem.  However,
important information at non-zero chemical potentials may be obtained using the well
established numerical technique of Taylor expansion of the QCD partition function
with respect to the baryon chemical potential \cite{taylor-exp}.  In this method one
Taylor expands the partition function ($\mathcal{Z}$)/free energy ($f$)/pressure ($P$)
in power series of the $\mu_B$ around $\mu_B=0$--- $P(\mu_B,T)/T^4 =
\sum_{n=0}^{\infty} (1/n!) \chi_n^B(T) (\mu_B/T)^n$, where $VT^3 \chi_n^B(T) = [
\partial^n \mathcal{Z} (\mu_B,T) / \partial (\mu_B/T)^n ]_{\mu_B=0}$ and $V$ is the
volume of the system. $\chi_n^B$ are the generalized baryon number susceptibilities
and since they are defined at $\mu_B=0$ usual lattice QCD simulations at zero
chemical potential can be used to compute these susceptibilities.  Similar
susceptibilities $\chi_n^Q$, $\chi_n^S$ \etc can be defined from the Taylor
expansions of the pressure with respect to the electric charge chemical potential
$\mu_Q$ , the strangeness chemical potential $\mu_S$ \etc.

\bfig
\subfigure[]{ \label{fig:chi2B} \includegraphics[scale=0.35]{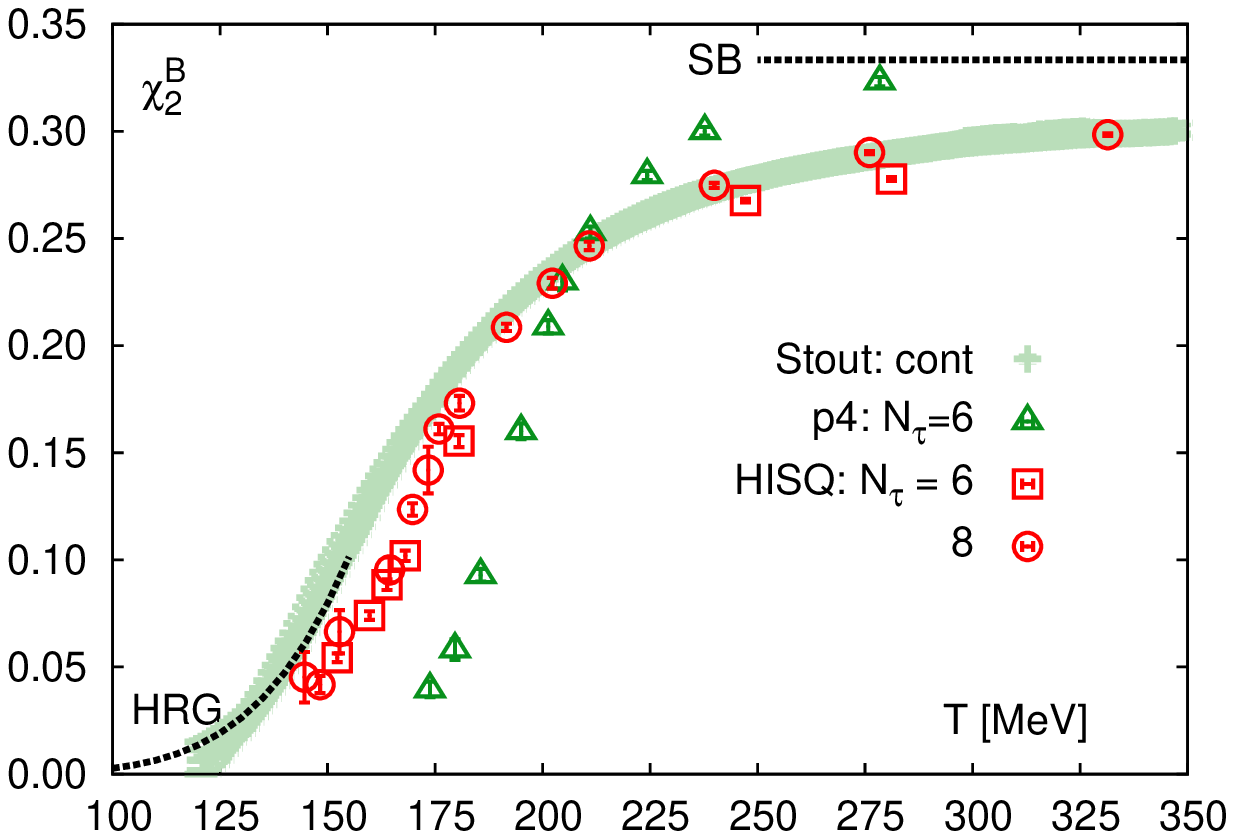} }
\subfigure[]{ \label{fig:chi4u} \includegraphics[scale=0.35]{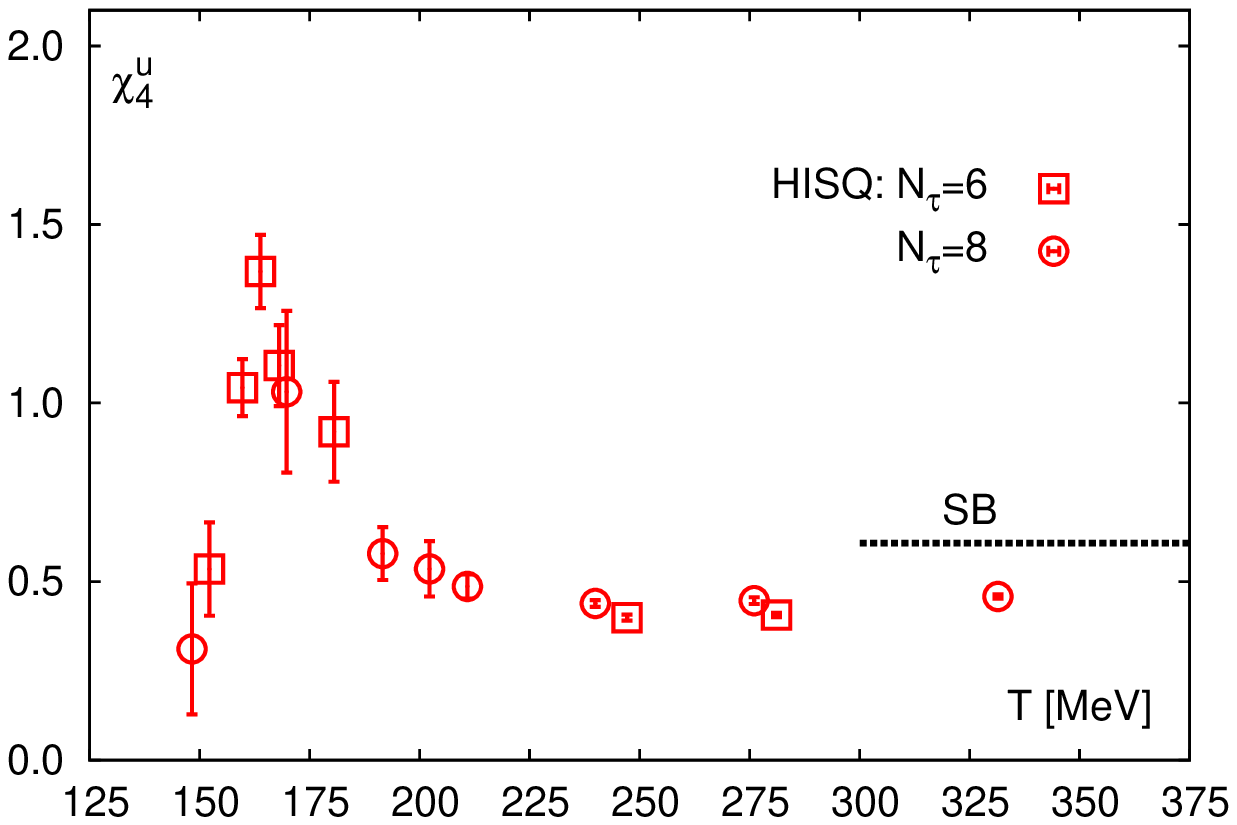} }
\subfigure[]{ \label{fig:chi6Q} \includegraphics[scale=0.35]{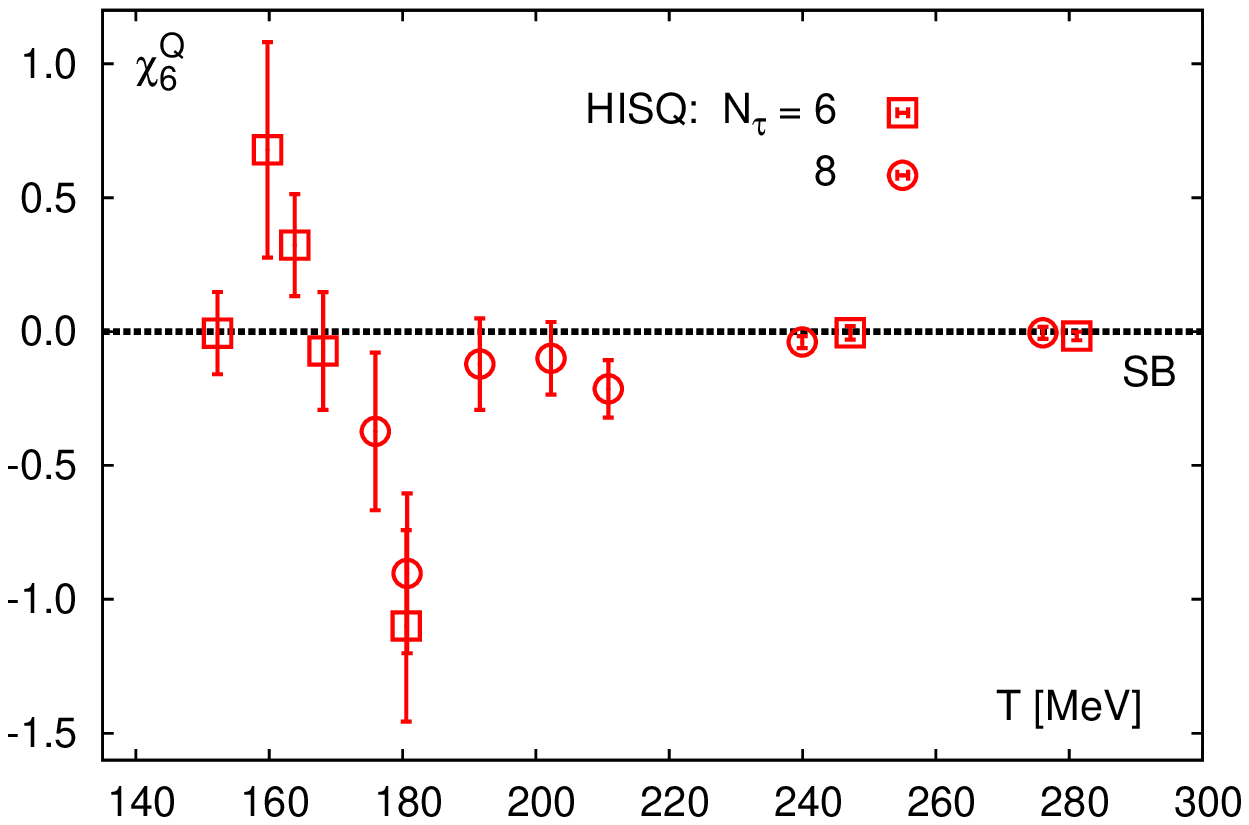} }
\vspace{-0.25cm}
\caption{(a) Second order susceptibility for the baryon number, (b) fourth order
susceptibility for up quark number and (c) sixth order susceptibility for the
electric charge. The new calculations for the HISQ action have been performed by the
BNL-Bielefeld collaboration and that for the stout action are from the
Wuppertal-Budapest collaboration \cite{ratti-1}. The older data for the p4fat3
action is taken from Ref.\ \cite{cheng-1}. The dotted lines indicate the values of
these quantities for the hadron resonance gas (HRG) model and for the free fermionic
gas (SB).} 
\efig

Since the generalized susceptibilities are derivatives of the free energy with
respect to the chemical potentials universal scaling properties of these
susceptibilities can also be readily predicted. From discussions in Sections
\ref{sec:scaling} and \ref{sec:curvature} it can be inferred that two derivatives of
the free energy with respect to $\mu_B$ is equivalent to a single derivative with
respect $T$ and consequently the scaling behavior of the susceptibilities are---
$\chi_n^B\sim h^{(1-\alpha-n/2)} f^{(n/2)}_s(z) + C_{n,reg}$. Here $f^{(n/2)}(z)$ are
unique scaling functions and $C_{n,reg}$ are contributions from the non-critical
regular part of the free-energy. In order to understand the universal scaling
behavior of the generalized susceptibilities it is also important to note that for
$3$-d $O(4)$ universality class the critical exponent $-1<\alpha<0$. In
\fig{fig:chi2B} we show the second order baryon number susceptibility.  Close to
$T_c$, $\chi_2\sim C_{2,reg}\mp A_{\pm}\ltvert(T-T_c)/T_c\rtvert^{(1-\alpha)}$ is
non-divergent and  behaves like the energy density. While for $T<T_c$ the new lattice
data is in good agreement with the hadron resonance gas model, for $T>T_c$ approach
towards the ideal gas value is much smoother and slower compared to the older data
from the p4fat3 action \cite{cheng-1}. Also note the good agreement between the new
data coming from the HISQ (BNL-Bielefeld collaboration) and the stout \cite{ratti-1}
actions. Close to $T_c$ the behaviors of the higher order susceptibilities are also
consistent with there expected scaling behaviors. The fourth order susceptibility
(see \fig{fig:chi4u}) $\chi_4\sim C_{4,reg}\pm
A_{\pm}\ltvert(T-T_c)/T_c\rtvert^{-\alpha}$ remains non-divergent and behaves like
the specific heat by always staying positive and showing a cusp like structure in the
vicinity of $T_c$. The sixth order susceptibility $\chi_6\sim\mp A_{\pm}
\ltvert(T-T_c)/T_c\rtvert^{-(1+\alpha)}$ changes sign across $T_c$ (see
\fig{fig:chi6Q}) and diverges at $T_c$ in the chiral limit. Based on the later
observation that even for $\mu_B\approx0$ $\chi_6<0$ in the vicinity of $T_c$, it has
been recently argued \cite{neg-chi6} that if the measured values of the sixth order
moments of the conserved charges come out to be negative for the on-going heavy-ion
collision experiments at the Large Hadron Collider (LHC) then it may indicate that
the transition from the hadronic to the QGP phase happens very close to the
freeze-out curve. 

The off-diagonal susceptibilities $\chi_{11}^{BS}$, $\chi_{11}^{QS}$ \etc, are
measures of correlations among different conserved charges. As a result these
quantities are good probes for the degrees of freedom inside QGP.  As for example,
inside QGP where strange quarks ($B=1/3$, $S=-1$) are the relevant degrees of freedom
strangeness can only exists in direct conjunction with the baryon number. On the
other hand, at very low temperatures in the hadronic phase where the relevant
strangeness carrying excitation are the kaons ($B=0$) there is no apparent
correlation among the baryon number and strangeness. Based on these observations the
observable baryon strangeness correlation $C_{BS}=-3\chi_{11}^{BS}/\chi_2^S$ was
proposed in Ref.\ \cite{koch-1}. In \fig{fig:cBS} we show the this quantity computed
using the HISQ (by BNL-Bielefeld collaboration) as well as using the stout
\cite{ratti-1} action. Below $T_c$ this quantity is in very good agreement with the
results predicted by the HRG model indicating that at these temperatures the
strangeness is truly carried by the hadrons. On the other hand, for $T\gtrsim250$ MeV
the baryon strangeness correlation is quite close to its ideal gas value indicating
that above these temperatures the strangeness and baryon number are carried by the
strange quarks. Also note that the lattice data from both the actions are in
excellent agreement with each other.

\bfig
\subfigure[]{ \label{fig:cBS} \includegraphics[scale=0.35]{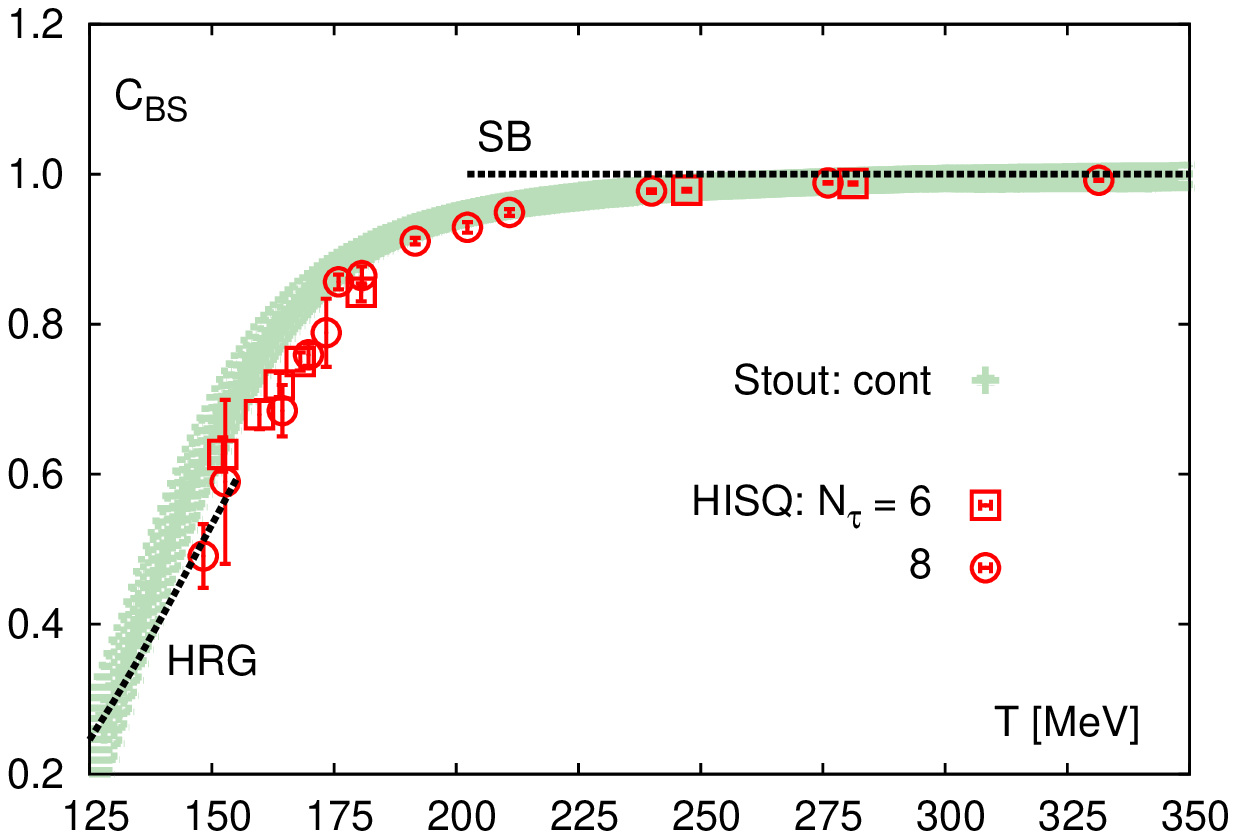} }
\subfigure[]{ \label{fig:star-lgt} \includegraphics[scale=0.35]{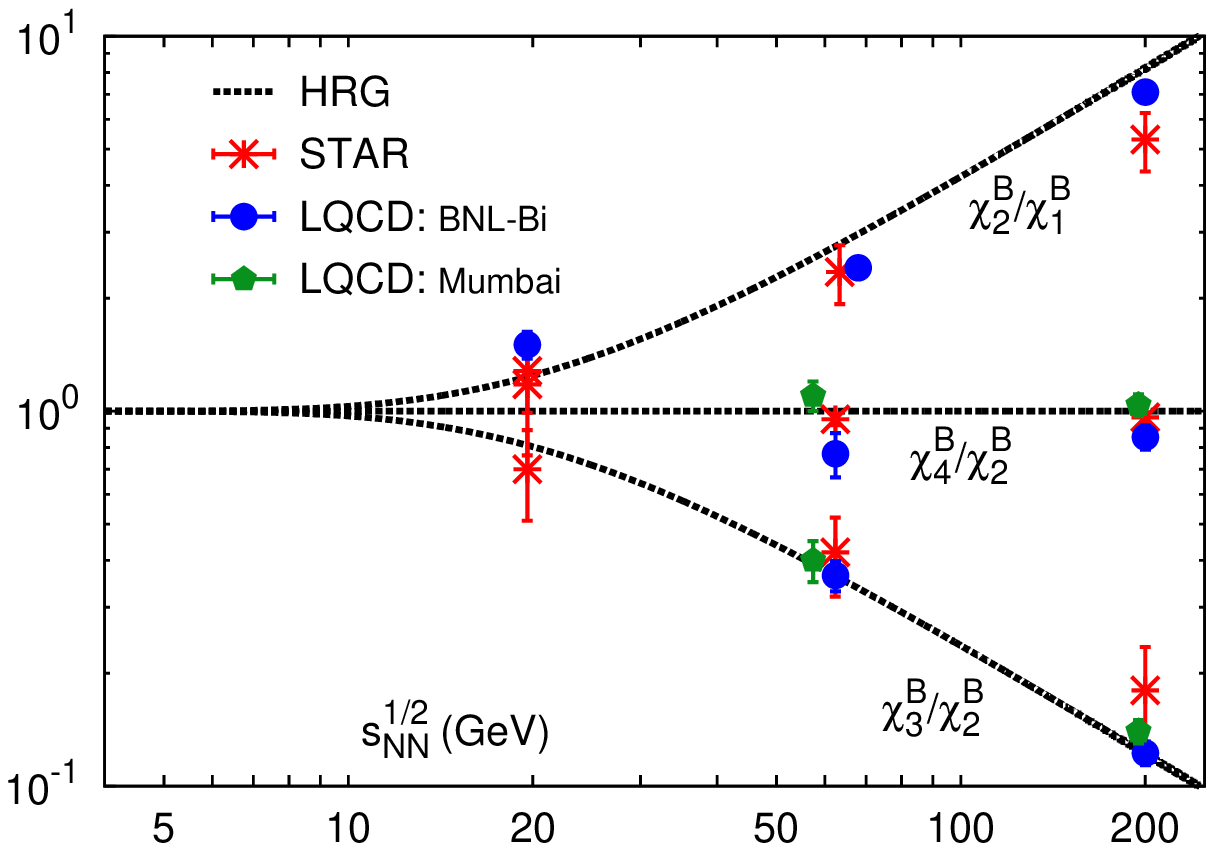} }
\subfigure[]{ \label{fig:diff-ord} \includegraphics[scale=0.35]{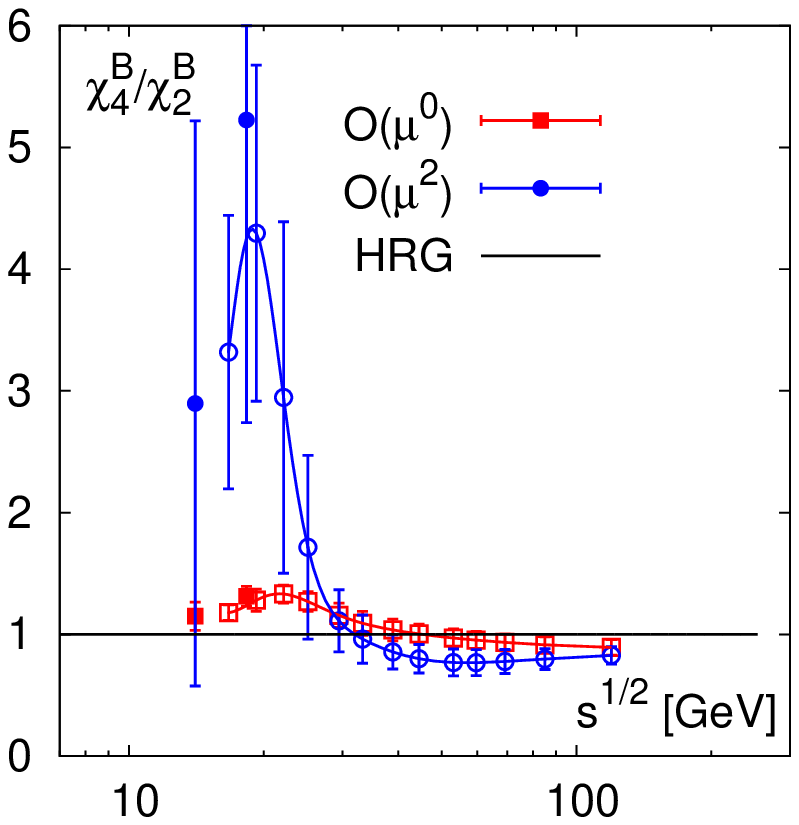} }
\vspace{-0.25cm}
\caption{(a) The baryon-strangeness correlation \cite{koch-1} using the HISQ
(BNL-Bielefeld) and the stout \cite{ratti-1} actions. The dotted lines indicate the
results for the hadron resonance gas model (HRG) and for the free fermionic gas (SB).
(b) Experimentally measured moments of net-proton fluctuations \cite{star-1} are
compared to lattice QCD computations of ratios of baryon number susceptibilities by
the BNL-Bielefeld \cite{schmidt-1} and the Mumbai \cite{gavai-1} groups. The HRG
results are from Ref.\ \cite{karsch-2}. (c) An example of the sensitivity of the
ratios of conserved charge susceptibilities to different orders of Taylor expansion
\cite{schmidt-1}. See text for detail.}  
\efig

So far we discussed the physics of the generalized susceptibilities at zero chemical
potentials. However, susceptibilities for non-zero values of chemical potentials have
more direct relevance for the heavy-ion collision experiments. These can be obtained
from the Taylor series expansions of the susceptibilities themselves, \eg the baryon
number susceptibilities at $\mu_B>0$ are given by--- $\chi_n^B(\mu_B,T) =
\sum_{k=0}^{\infty} (1/k!) \chi_{n+k}^B(0,T) (\mu_B/T)^k$. These generalized
susceptibilities are measures of the moments of the conserved charge distributions.
As for example--- $VT^3\chi_1^B(\mu_B,T)=\lang N_B\rang$,
$VT^3\chi_2^B(\mu_B,T)=\lang(\delta N_B)^2\rang$, $VT^3\chi_3^B(\mu_B,T)=\lang(\delta
N_B)^3\rang$, $VT^3\chi_4^B(\mu_B,T)=\lang(\delta N_B)^4\rang-3\lang(\delta
N_B)^2\rang^2$ \etc measure different moments of the distribution of the net baryon
number $N_B(\mu_B,T)$, where $\delta N_B(\mu_B,T)=N_B(\mu_B,T)-\lang
N_B(\mu_B,T)\rang$. On the other hand, such moments of conserved charges have also
been measured by the STAR experiment \cite{star-1} using event-by-event fluctuations
in the Relativistic Heavy Ion Collider (RHIC) at various energies ($\sqrt{s_{NN}}$) .
As for example, in heavy-ion collision experiments one measures the moments--- mean
$M_B(\sqrt{s_{NN}})=\lang N_B\rang$, variance $\sigma_B^2(\sqrt{s_{NN}})=\lang(\delta
N_B)^2\rang$, skewness $S_B(\sqrt{s_{NN}})=\lang(\delta N_B)^3\rang/\sigma_B^3$,
kurtosis $K_B(\sqrt{s_{NN}})=\lang(\delta N_B)^4\rang/\sigma_B^4-3$ \etc.
Assuming--- (i) experiments measure conserved charge distributions of a thermalized
system, (ii) measured moments characterize the chemical freeze-out condition and
(iii) $T(\sqrt{s_{NN}})$ and $\mu_B(\sqrt{s_{NN}})$ at the chemical freeze-out can be
modeled by the hadron resonance gas model \cite{cleymans-1}, the experimentally
measured volume independent combinations of the moments can be related to the ratios
of susceptibilities computed from lattice QCD---
$\sigma_B^2(\sqrt{s_{NN}})/M_B(\sqrt{s_{NN}})=\chi_2^B(\mu_B,T)/\chi_1^B(\mu_B,T)$,
$\sigma_B(\sqrt{s_{NN}})S_B(\sqrt{s_{NN}})=\chi_3^B(\mu_B,T)/\chi_2^B(\mu_B,T)$,
$\sigma_B^2(\sqrt{s_{NN}})K_B(\sqrt{s_{NN}})=\chi_4^B(\mu_B,T)/\chi_2^B(\mu_B,T)$
\etc.  

As net proton fluctuations can be treated as a proxy for the net baryon number
fluctuations \cite{hatta-1}, in \fig{fig:star-lgt} we show a comparison of the
experimentally measured moments of net proton distribution \cite{star-1} to the
lattice QCD calculations of the ratios of the baryon number susceptibilities using
improved p4fat3 \cite{schmidt-1} and un-improved naive \cite{gavai-1} staggered
actions. As can been seen, the experimental data and the lattice QCD computations are
in good agreement with each other and also with that from the hadron resonance gas
calculations \cite{karsch-2}. This, for the first time, shows a direct comparison
between the heavy-ion collision experiments and lattice QCD computations. However, in
order to make such comparisons it is essential to perform reliable computations of
the higher order susceptibilities. As for example, in \fig{fig:diff-ord} we show the
results \cite{schmidt-1} for ratio $\chi_4^B/\chi_2^B$ computed in leading and next
to leading orders of the Taylor expansion. As can be seen, around
$\sqrt{s_{NN}}\sim20$ GeV (\ie $\mu_B\sim200$ MeV) the contribution of the next to
leading order term is much larger than the contribution of the leading order term.
Moreover, at present these lattice results are not continuum extrapolated and were
calculated using actions which are known to suffer from significant discretization
effects and resulting in larger values for the transition temperature. Hence, in
future, it is essential to carry out these lattice computations using HISQ like
highly improved actions with accurate determination of the higher order
susceptibilities. This is specially important  for the observables like moments of
net electric charge fluctuations \cite{star-1} as they are sensitive to light charged
hadrons like the pions (see discussion in Section \ref{sec:Tc}). Such calculations
are currently in progress


\vspace{0.5cm}\noindent
{\sl Acknowledgments}: The author is supported under Contract No.  DE-AC02-98CH10886
with the U.S.  Department of Energy.


\end{document}